\documentclass[a4paper,10pt]{article} %
\usepackage{graphicx,amssymb} %

\date{} %

\def\keywords#1{\begin{center}{\bf Keywords}\\{#1}\end{center}} %

\def\titulo#1{\title{#1}} %
\def\autores#1{\author{#1}} %

\newcounter{note}
\usepackage{hyperref}

\usepackage{amssymb}
\usepackage{amsmath}
\usepackage{dsfont} 
\usepackage{listings} 
\usepackage{courier} 

\newcommand{\eg}{\emph{e.g}}
\newcommand{\ie}{\emph{i.e.}}

\newcommand{\Mathematica}{\emph{Mathematica}}
\newcommand{\MMA}{\Mathematica}

\DeclareMathOperator{\res}{\mathbf{res}}
\DeclareMathOperator{\unres}{\mathbf{unres}}
\DeclareMathOperator{\tr}{\mathbf{tr}}

\newcommand{\C}{\ensuremath{\mathds{C}}}
\newcommand{\Cplx}{\ensuremath{\C}}

\newcommand{\Jamiolkowski}[1]{\ensuremath{\mathcal{J}_{#1}}}

\newcommand{\MatrixForm}[1]{\ensuremath{\mathcal{M}_{#1}}}
\newcommand{\NatRep}[1]{\MatrixForm{#1}}
\newcommand{\GenNatRep}[2]{{\MatrixForm{#1}^{#2}}}
\newcommand{\M}{\ensuremath{\mathds{M}}}
\newcommand{\Id}{\ensuremath{\mathds{1}}}
\newcommand{\SWAP}{\ensuremath{\mathrm{SWAP}}}

\newtheorem{definition}{Definition}

\begin{document}

\titulo{Functional framework for representing and transforming quantum channels}

\autores{Jaros{\l}aw Adam Miszczak\\ %
       Institute of Theoretical and Applied Informatics,\\ 
       Polish Academy of Sciences\\ 
       Ba{\l}tycka 5, 44-100 Gliwice, Poland\\ \\
       \tt{miszczak@iitis.pl}
       }%

\maketitle

\thispagestyle{empty}


\begin{abstract}
We develop a~framework which aims to simplify the analysis of quantum states and
quantum operations by harnessing the potential of function programming paradigm.
We show that the introduced framework allows a seamless manipulation of
quantum channels, in particular to convert between different representations of
quantum channels, and thus that the use of functional programming concepts
facilitates the manipulation of abstract objects used in the language of quantum
theory.

For the purpose of our presentation we will use \emph{Mathematica} computer
algebra system. This choice is motivated twofold. First, it offers a rich
programming language based on the functional paradigm. Second, this programming
language is combined with powerful symbolic and numeric manipulation
capabilities.
\end{abstract}


\section{Introduction}

Functional programming is frequently seen as an attractive alternative to the
traditional methods used in scientific computing, which are based mainly on the
imperative programming paradigm~\cite{hinsen09promises}. Among the features of
functional languages which make them suitable for the use in this area is the
easiness of execution of the functional code in the parallel environments.

The main aim of this work is to show that the functional programming
concepts facilitate the use of abstract objects used in the language of
quantum theory. We develop a~framework which aims to simplify the analysis of
quantum states and quantum operations by harnessing the potential of functional
programming paradigm. For the purpose of our presentation we will use
\MMA\ computer algebra system. This choice is motivated twofold. First,
it offers a rich programming language based on the functional paradigm. Second,
this programming language is combined with powerful symbolic and numeric
manipulation capabilities.

During the last few years a number of simulators of quantum information processing
has been developed using \MMA\ computing system~\cite{touchette00qucalc,
juliadiaz06qdensity, tabakin11qcwave, quantum2}.
Unfortunately, these packages do not use functional programming capabilities of
this system and are focused on pure states and unitary operations. Moreover, they
focus on the quantum mechanical systems which can be represented using state
vectors and include only a basic functionality required for the purpose of
manipulating and analyzing quantum states.

In this paper we follow the pragmatic approach and we provide a set of useful
constructions which can be helpful for the analysis of quantum channels. At the
same time we advocate the use of functional programming in this approach. We
argue that by using the functional language elements provided by \MMA\ one can
easily and efficiently convert between different representations of quantum
channels.

\section{Functional syntax for quantum channels}\label{sec:channels-functional}

\subsection{Notation}\label{sec:intro}
In the following we assume that the quantum systems are represented by
finite-dimensional density matrices, \ie\ positive semidefinite complex matrices
with unit trace. The space of density matrices of
dimension $d$ is denoted by $\Omega_d$. We use $\res$, $\unres$ operations
\cite{miszczak11singular} for converting between matrix and vector forms of
states and operators. In the \MMA\ language function $\res$ is defined as a
synonym for a built-in function
\lstinline+Flatten+
\begin{lstlisting}
Res = Function[m, Flatten[m]];
\end{lstlisting}
This function transforms a matrix \lstinline+m+ into a vector in a row order.
Function $\unres$, which is a reverse transformation, is defined in \MMA\
as
\begin{lstlisting}
Unres = Function[m, Partition[m, Sqrt[Length[m]]]];
\end{lstlisting}
and it uses built-in function \lstinline+Partition+ to get back from a
one-dimensional list to a matrix. As the space of density matrices is unitary with
Hilbert-Schmidt scalar product, we introduce a function
\begin{lstlisting}
HSInner = Function[x, Function[y, Tr[x.ConjugateTranspose[y]]]];
\end{lstlisting}
which, thanks to the curried form, allows using a partial application in the
application of this scalar product.

Unfortunately \MMA\ does not provide a straightforward support for the partial
application of functions. The language does not allow using functions with too
few parameters and one has to explicitly use empty slots
(\lstinline{#}-signs) to define a partially applied function. For this reason in
order to use the functional version of some procedures, it is necessary to provide a
curried version of these functions.

\subsection{Simple channels}
Let us illustrate the above considerations with the simplest example -- the
transposition map. This map is defined as
\begin{equation}
\rho\mapsto \rho^T,
\end{equation}
and can be expressed in \MMA\ as
\begin{lstlisting}
trans = Function[x, Transpose[x]]
\end{lstlisting}
or using more compact syntax as \lstinline{trans = Transpose[#]&}.

One should note that this map is not completely positive, hence it does not
represent a valid quantum channel. Nevertheless, it is useful for presenting
basic transformations which can be performed on quantum channels.

If we would like to apply this function on some state $\rho$ we simply write
\lstinline{trans[$\rho$]}. In many situations however, one needs to
apply a map on a list \lstinline{rhos} of states or matrices. In this case we
simply map the functions representing the map on the list using \lstinline+Map+
function as \begin{lstlisting}
Map[trans, rhos]
\end{lstlisting}
or using more compact syntax as \lstinline+trans /@ rhos+.

\subsection{Channels with parameters}
In order to use channels defined by parametrized expression, one can employ
partially applied functions. The simplest example of such channels is a
depolarizing channel $\Psi_{D(p,d)}$ defined as
\begin{equation}\label{eqn:depolar-channel}
\Psi_{D(p,d)}(\rho) = (1-p) \rho + p \frac{1}{n}\Id_n,
\end{equation}
where we assume that $\rho\in\M_n$ and $\Id_n$ denotes the identity matrix of
the appropriate size. 

Using the notation introduced in Section \ref{sec:intro}, this channel can be
represented by a function
\begin{lstlisting}
dep = Function[d, Function[p, Function[x, 
    (1-p)x + p IdentityMatrix[d]/d
]]];
\end{lstlisting}
Here we follow the convention that the function parameters should be organized
in such a way, that by providing all but one of them, we obtain a function
accepting quantum state as an argument. In the above case the first two
parameters represent the dimension and the reliability of the channel
(the probability of introducing no errors).

Function \lstinline{dep} requires three arguments and its application on state
$\rho$ is achieved by first declaring the instance of the channel for a fixed
dimension (\eg\ d=4)
\begin{lstlisting}
dep4 = dep[4];
\end{lstlisting}
and next using this function with a specific probability \lstinline{p}
\begin{lstlisting}
dep4[p][$\rho$];
\end{lstlisting}

However, one can use \lstinline{dep} function to define the expression in which
only two arguments are provided
\begin{lstlisting}
g = (dep[#1][p][#2]) &
\end{lstlisting}
and this allows obtaining a general definition of the depolarizing channel with
a fixed parameter \lstinline{p}, identical to the following definition
\begin{lstlisting}
Function[d, Function[x, (1-p) x + p IdentityMatrix[d]/d]];
\end{lstlisting}

Function \lstinline{g} accepts two arguments representing the dimension and the
input state. Its application on some state $\rho\in\M_4$ reads
\begin{lstlisting}
g[4][$\rho$];
\end{lstlisting}
and this syntax allows the selection of an argument which should be fixed during the
manipulation.

\section{Representations of quantum channels}\label{sec:channels-representations}

\subsection{Natural representation}\label{sec:natural-rep}
As channels are linear mappings, it is possible, at least in finite-dimensional
case, to represent them by matrices. Let us assume that we are dealing with
$d=n\times n$ dimensional matrices. 

The base in $n^2$-dimensional space $\M_n$ is given by matrices,
which can be obtained by using \lstinline{Unres} operations on the base vectors
in the $d$-dimensional space $\Cplx^d, d=n^2$, as
\begin{lstlisting}
base = Map[Unres[UnitVector[d, #]] &, Range[d]];
\end{lstlisting}
where \lstinline{Range[d]} returns a list containing numbers $1,2,\dots,d$. In
the following we assume that the $d$-dimensional matrix base can be obtained
using function \lstinline{BaseMatrices[d]} defined as
\begin{lstlisting}
BaseMatrices =  Function[d, Map[Unres[UnitVector[d, #]] &, Range[d]]];
\end{lstlisting}

If the list \lstinline{fBase} contains the images of the quantum channel
\lstinline{f} on the base
\begin{lstlisting}
fBase = f /@ base
\end{lstlisting}
then the natural representation can be calculated by unreshaping the images of
the map on the base matrices in $\M_{d^2}$,
\begin{lstlisting}
{Res /@ fBase}
\end{lstlisting}
Combining this into one function gives
\begin{lstlisting}
NaturalRepresentation = Function[f, Function[d, 
    With[{base=BaseMatrices[d^2]}, Map[Res[f[#]]&, base]]
]; 
\end{lstlisting}

We denote the natural representation of the channel $\Phi$ by
$\MatrixForm{\Phi}$, assuming that this matrix is obtained in the standard
basis. Matrix $\MatrixForm{\Phi}$ is sometimes called a supermatrix for the
channel $\Phi$.

The above considerations can be summarized as the following definition.
\begin{definition}[Natural representation]\label{def:natural-rep}
For a given channel $\Psi$, the natural representation of $\Phi$ by
$\NatRep{\Phi}$ is defined as
\begin{equation}
(\NatRep{\Phi})_{i.} = \res \Phi ( b_{i})
\end{equation}
where $(A)_{i.}$ denotes $i$-th column of the matrix $A$ and $b_i$, $i=1,n^2$
denotes base matrices in $\M_n$.
\end{definition}

For example, in order to obtain the matrix representation of the depolarizing
channel \lstinline{dep} acting on one qubit, one should use
\lstinline{NaturalRepresentation} function as
\begin{lstlisting}
NaturalRepresentation[dep[2][p]][2]
\end{lstlisting}
In a similar manner one can check that the natural representation of the
one-qubit transposition channel \lstinline{trans}
\begin{lstlisting}
NaturalRepresentation[trans][2]
\end{lstlisting}
is equal to the \SWAP\ gate.

\subsection{General natural representation}
Clearly one can represent a given channel in a matrix form using not only a
canonical base, but any orthonormal basis in $\Cplx^{n^2}$. In this situation
one cannot use the method described above as it relies on the special form of
the canonical base matrices.

The straightforward method of calculating a matrix representation, is based
on the formula
\begin{equation}
(M_\Phi^b)_{ij}=\tr [ b_i\Phi(b_j)^\dagger ],
\end{equation}
where $b_i,i=1,\dots,n^2$ denotes the base. 

\begin{definition}[General natural representation]\label{def:general-natural-rep}
For a given channel $\Psi$, the general natural representation of $\Phi$ in base
$b$ is defined as
\begin{equation}
(\GenNatRep{\Phi}{b})_{ij} = \tr[ \Phi (b_{i}) b_{j}^\dagger],
\end{equation}
where $b_i$, $i=1,n^2$ denote base matrices in $\M_n$.
\end{definition}

This definition can be implemented using \lstinline/Outer/ function as
\begin{lstlisting}
Function[f, Function[b, 
    Outer[HSInner[#1][#2]&, Map[f,b], b, 1]
]];
\end{lstlisting}
where \lstinline/base/ is a given base or, alternatively, by using \lstinline/Map/ function as
\begin{lstlisting}
Function[f, Function[b, 
    Map[Map[#, b] &, Map[HSInner, Map[f, b]]]
]];
\end{lstlisting}

This method
requires $n^4$ multiplications of $n\times n$ matrices and is highly
inefficient.

The simplest method is to reconstruct a change of basis matrix $M_B$,
\begin{lstlisting}
$M_B$ = Map[Res, b]
\end{lstlisting}
and use it to obtain $M_\Phi^b$ as
\begin{equation}
M_\Phi^b = M_B^{\phantom\dagger } M_\Phi^{\phantom b } M_B^\dagger.
\end{equation}

\subsection{Choi-Jamio{\l}kowski representation}
Complete positivity, one of the
requirements for the map between finite-dimensional spaces can be formulated
using Choi-Jamiolkowski representation of a map~\cite{jamiolkowski72linear,
choi75completely}. This representation in the context of quantum channels is
known as Jamiolkowski isomorphism and here the image of this isomorphism is
denoted as $\Jamiolkowski{\Phi}$.

The Choi-Jamiolkowski representation is closely related to the natural
representation. The natural representation of the channel acting on $n\times
n$-dimensional matrices is always obtained with respect to some bases
$\{b_i\}_{i=1,n^2}$, where $n^2$ is the dimension of the state space. 

If one uses base $b_i$ to obtain the natural representation of the channel $\Phi$
resulting in matrix $M_{\Phi}^b$, then the Choi-Jamiolkowski matrix for this
channel is obtained as 
\begin{equation}
\{\Jamiolkowski{\Phi}^b\}_{i,j} = \tr[ M_{\Phi}^b (b_i\otimes b_j)],
\end{equation}
for $i,j=1,n^2$.

\begin{definition}[Choi-Jamiolkowski matrix]
Let $\{b_i\}$ be a base in $\Cplx^{n^2}$. The Choi-Jamiolkowski matrix
corresponding to a general natural representation in base $\{b_i\}$ is defined
as
\begin{equation}
\{\Jamiolkowski{\Phi}^b\}_{i,j} = \tr[ M_{\Phi}^b (b_i\otimes b_j)].
\end{equation}
\end{definition}

The Choi-Jamiolkowski representation of the channel $\Phi$ can be also obtained
using several other methods. One of the simplest formulas is the one expressing
$\Jamiolkowski{\Phi}$ as a sum
\begin{equation}
\Jamiolkowski{\Phi} = \sum_{i=1}^d \Phi(e_i)\otimes e_i.
\end{equation}
Assuming that \lstinline{base} represents matrix base in $d$-dimensional space,
this representation can be used by mapping 
\begin{lstlisting}
cjBase = Map[KroneckerProduct[f[#], #] &, base]
\end{lstlisting}
and accumulating the results
\begin{lstlisting}
Plus /@ cjBase
\end{lstlisting}
Combining the above into one function gives
\begin{lstlisting}
ChoiJamiolkowskiRepresentation = Function[f, Function[d, 
    With[{base=BaseMatrices[d]}, 
        Map[Plus,[Map[KroneckerProduct[f[#],#]&, base]]]
]];
\end{lstlisting}

The Choi-Jamiolkowski representation of a channel is related to the natural
representation, one can easily construct a Choi-Jamiolkowski matrix
corresponding to a given generalized natural representation.

\paragraph*{Acknowledgements}
This work was supported by the Polish Ministry of Science and Higher Education
under the grant number IP2011 036371 and by the Polish National Science Centre
under the grant number DEC-2011/03/D/ST6/00413. Author would like to acknowledge
stimulating discussions with Z.~Pucha{\l}a, P.~Gawron, D.~Kurzyk, V. Jagadish
and P. Zawadzki.

\bibliographystyle{plain}
\bibliography{functional_channels}

\end{document}